# Understanding and Evaluating Engineering Creativity: Development and Validation of the Engineering Creativity Assessment Tool (ECAT)

Zeynep G Akdemir-Beveridge[1,*], Arash Zaghi[1], and Connie Syharat[1]
[1]College of Engineering, University of Connecticut, Storrs, CT 06269

*Corresponding author, zeynep.akdemir@uconn.edu

## Abstract

Creativity is essential in engineering education, helping students develop innovative, practical solutions. Yet, assessing it remains challenging due to a lack of reliable, domain-specific tools. Traditional assessments, such as the Torrance Tests of Creative Thinking (TTCT), may not fully capture the multifaceted nature of engineering creativity. This study introduces and validates the Engineering Creativity Assessment Tool (ECAT) as an alternative tailored to engineering students. We evaluated the validity and reliability of ECAT using data from 199 undergraduate engineering students who completed a hands-on creative design task. Five trained raters evaluated student products using the ECAT rubric. Exploratory and Confirmatory Factor Analyses (EFA and CFA) were conducted to examine ECAT's structure. Reliability was assessed via Cronbach's alpha (internal consistency) and Kendall's Tau-b (inter-rater agreement), while convergent and discriminant validity were examined using TTCT scores. Factor analyses supported a four-factor model (fluency, originality, cognitive flexibility, and creative strengths) with acceptable fit (CFI = 0.95, TLI = 0.93, RMSEA = 0.089). Internal consistency was high ($\alpha = 0.89$), and inter-rater reliability ranged from 0.71 to 0.84 ($p < 0.01$). ECAT showed moderate convergent validity with TTCT ($r = 0.26$–$0.32$), supporting its utility in capturing domain-specific creativity. ECAT provides a reliable and valid framework for assessing creativity in engineering contexts. Its multi-dimensional design offers actionable insights for educators aiming to foster creativity in students. Future research should examine ECAT's applicability across varied student populations, course types, and instructional settings.

## Introduction

In an era where technological advancements are transforming every aspect of our lives, the demand for creative engineering solutions has never been higher. Consider the rapid development of electric vehicles as a response to the global climate crisis. Engineers are not only tasked with improving battery efficiency but also with reimagining the entire transportation ecosystem to reduce carbon emissions. Similarly, during the COVID-19 pandemic, engineering teams worldwide had to pivot quickly, designing ventilators from low-cost materials and developing rapid testing kits. These breakthroughs in real-life illustrate that solving such complex, urgent problems require more than technical know-how; it necessitates creativity, that is the ability to envision unconventional solutions, adapt them to real-world constraints, and implement them effectively (Cropley, 2015a).

Companies in fields like aerospace, renewable energy, and biotechnology require engineering graduates who can not only solve technical problems but also predict future needs and create disruptive technologies. For example, the decline of companies like Kodak and Nokia has often been attributed to a failure to innovate, as they were unable to anticipate or respond creatively to the rapid changes in technology and market demands (Lucas & Goh, 2009). Despite the critical need of promoting creativity and innovation, traditional engineering education has often emphasized analytical problem-solving and technical skills at the expense of nurturing creativity (Cropley, 2015b; Daly et al., 2014). Consequently, a lack of knowledge of creativity in engineering education can lead to stagnation in technology development, where incremental improvements replace groundbreaking innovations.

Engineering students often feel constrained by rigid curricula that prioritize standardized solutions over exploration and experimentation (e.g., Kazerounian & Foley, 2007). This can hinder their ability to approach problems from multiple perspectives and to think critically about alternative solutions. Similarly, Cropley (2015a) stated that contemporary engineering education tends to produce "convergent thinkers" who are skilled at applying existing knowledge to solve well-defined problems but often struggle with "divergent thinking," which is essential for generating novel solutions and tackling ambiguous, real-world challenges. Cropley argued that this overemphasis on technical proficiency and standardized testing leaves little room for developing the creativity and innovative thinking necessary to address the complex, ill-structured problems engineering graduates face in practice. Understanding this will help implement targeted strategies to effectively cultivate creativity as well as identifying teaching and learning gaps. This is where we evidently need comprehensive assessment tools for engineering creativity (Cropley, 2015b).

Traditional metrics often focus narrowly on innovation and problem-solving skills, but these alone do not capture the full spectrum of creativity that modern engineers need (Daly et al., 2014). While these assessments can measure students' technical skills and their ability to innovate within given constraints, they often overlook other critical aspects of creativity, such as encouraging tolerance of ambiguity, pushing students to reach their full potential by delaying gratification, or challenge conventional thinking (Cropley, 2015b; Hu et al., 2016). Therefore, this study aims to fill this gap by evaluating the creative productivity of engineering students through a comprehensive reliable and valid creativity assessment tool tailored to the field of engineering education. To achieve this, we develop the Engineering Creativity Assessment Tool (ECAT) and investigate its validity and reliability in assessing engineering students' creativity. We hypothesize that ECAT will reveal multiple distinct dimensions of creativity (mirroring divergent thinking components) and will correlate with, yet be distinct from, general creativity tests (i.e., TTCT), reflecting the domain-specific nature of engineering creativity.

## Motivation and Background

### *Fundamentals of Creativity*

Creativity is broadly understood as the ability to generate novel and appropriate solutions, products, or ideas in response to open-ended tasks or problems (Amabile, 2012). It involves divergent thinking, characterized by fluency, originality, flexibility, and elaboration that are the dimensions first identified by Guilford (1957). Contemporary creativity research emphasizes that creativity is a complex, multi-dimensional construct, going beyond mere novelty to integrate appropriateness and feasibility, particularly in applied fields such as engineering (Cropley & Cropley, 2010).

An influential framework for understanding creativity is the 4P model, conceptualizing creativity across four key dimensions: Person, Process, Product, and Press (environment). Initially proposed by Rhodes (1961), this model has been widely adopted and further expanded in contemporary creativity literature (Runco & Jaeger, 2012; Kaufman & Beghetto, 2013;

Glăveanu, 2013). The "Person" dimension emphasizes traits and skills that foster creativity, such as intrinsic motivation, cognitive flexibility, and domain expertise. The "Process" dimension highlights the cognitive steps individuals take in generating creative solutions. The "Product" dimension evaluates the outcomes of creative activities, assessing their originality, utility, and effectiveness. Lastly, the "Press" dimension refers to contextual and environmental factors influencing creative behaviors. Most existing creativity tests focus on the Product (outcome) aspect but often overlook Process and Press (contextual) factors, which are crucial in engineering design. This underscores the need for an assessment like ECAT that evaluates not only the final product's creativity but also aspects of the creative process (e.g., the flexibility and thought process behind the product).

Contemporary creativity research underlines that creativity is a multidimensional construct that varies from everyday creative acts to paradigm-shifting innovations capable of transforming entire fields or industries (Baer, 2017). Amabile's (2012) Componential Theory emphasizes that creativity depends on domain-specific expertise, creativity-relevant skills, and intrinsic task motivation. Furthermore, effective assessment of creativity requires domain familiarity from evaluators, underscoring the importance of context-specific criteria. For example, in the context of design and engineering, creativity plays a critical role in enabling innovation, driving the evolution of products, processes, and systems. Engineering creativity specifically demands a dual emphasis on novelty and usefulness, as the field inherently seeks practical, innovative solutions that are feasible and valuable within real-world contexts (Sarkar & Chakrabarti, 2011; Shah et al., 2003). In this context, Sarkar and Chakrabarti (2011) argue that assessing creativity in engineering involves understanding both novelty and usefulness, proposing methodologies to measure these aspects quantitatively. This dual focus is vital as creativity within engineering education is aimed not only at originality but also at addressing pragmatic needs effectively.

### *Creativity in Engineering Education*

Having outlined the general theory of creativity, we now focus on the specific context of engineering education, where creativity is increasingly recognized as crucial but faces unique challenges. There is a growing focus on fostering creativity among aspiring engineers, recognizing its crucial role in the engineering profession. National Academy of Engineering leaders have recently emphasized that the 'hallmark of engineering is creativity' – producing solutions we didn't know we needed but now cannot live without (Frueh, 2024). This sentiment echoes the consensus among engineering students that creativity is essential and should be more prominent in their educational curricula.

According to Cropley (2015b), a lack of educator training in creativity and an absence of a 'habit of creativity' in engineering programs remain significant issues. This leads to curricula that do not intentionally cultivate creative skills, even though students and educators acknowledge creativity's importance. To develop "a habit in mind" in any context, it is crucial to understand the psychological principles, which are often supported by theories from the learning sciences, psychology, and neuroscience (e.g., Serice, 2022; Zeine et al., 2024). The process of creative problem-solving in engineering has a comprehensive nature, which guides us in generating solutions to novel, challenging problems, including how to generate them, who can generate them, how to recognize them, and how to simulate them (Cropley, 2015b).

Creativity is recognized when there is "the production of a novel and appropriate response, product, or solution to an open-ended task." (Amabile, 2012 p.3). Amabile's Componential Theory of Creativity (Amabile, 2012) suggests that the level of creativity of a response, product, or solution depends on the extent that it is seen creative by people who are familiar with the domain. This means that creativity is a spectrum of skills or a gradient that can be categorized into different levels depending on specific functions and contexts. Basically, it is not suggested to call someone creative/uncreative on a dichotomic scale, rather we can find one to be more or less creative due to its continuous nature (Baer, 2017). These levels range from everyday personal creativity to groundbreaking innovations that can potentially transform the entire field. Therefore, one needs to consider the construct of creativity not only in terms of creativity-relevant skills such as divergent thinking (Guilford, 1957) but also in conjunction with domain-specific skills that provide *all forms of knowledge* in different levels (i.e., declarative/factual, procedural, conditional, functional) and *adaptive expertise* necessary to generate innovative solutions in the context of engineering (Cropley, 2015b).

Cropley (2015b) describes this issue as a "lack of knowledge" problem in connecting engineering with creativity, where engineering educators can be categorized into three groups. The first group lacks knowledge about creativity and how it can be taught in engineering contexts, resulting in minimal real progress. The second group focuses primarily on teaching factual or declarative knowledge, which prevents learners from reaching the threshold of adaptive expertise without incorporating creativity. The third group, on the other hand, emphasizes over-specialization in degrees, solely focusing on technical content, which also leaves no room for creativity despite achieving a high level of expertise in engineering. This lack of knowledge on how to incorporate effective creative engineering practices highlights the need to understand how to cultivate a habit of creative thinking in the field of engineering education.

### *Challenges in Evaluating Engineering Creativity*

Assessment and evaluation of creativity in engineering education poses significant challenges that stem from the inherent nature of engineering disciplines, the pedagogical approaches employed, lack of consistent assessment approaches, and the perceptions surrounding creativity itself. One of the primary challenges is the traditional focus on closed-ended problems within engineering curricula, which often prioritize algorithmic solutions over creative thinking. Core engineering courses typically present problems with a singular correct answer, thereby limiting opportunities for students to engage in creative problem-solving (Hirshfield & Koretsky, 2020). This deterministic learning environment can suffocate the development of creative skills, as students may not perceive creativity as a valuable component of their engineering learning (Carpenter, 2016).

Another major challenge is that many engineering educators are reluctant to integrate creativity into their teaching practices (Tekmen-Araci & Mann, 2018). This reluctance stems from factors such as a focus on product over process, a performance-driven mindset, risk aversion, and external pressures related to course structure and content delivery (Roncin, 2011). Additionally, a lack of knowledge on how to effectively teach and assess creativity in the engineering field contributes to resistance from both educators and institutional frameworks (Dorrington et al., 2019). This issue may be further compounded by students' perception that creativity is largely absent in engineering courses, leading to decreased engagement and motivation to explore creative opportunities in the field. A comprehensive creativity assessment tool (such as the proposed ECAT) would help instructors gain clear and practical criteria for identifying and assessing creativity. By explicitly evaluating certain engineering creativity dimensions, ECAT can provide a structured approach that reduces uncertainty around teaching creativity, support more open-ended assignments, and foster a learning space where exploration and innovation are encouraged.

Engineering educators face significant challenges in assessing how well their pedagogical strategies cultivate creativity among students, particularly due to the absence of a clear, domain-specific framework for evaluating creativity in engineering (Kazerounian & Foley, 2007). While instructors across different fields, such as project-based assessments in engineering, portfolio reviews in art, and standardized tests in creative writing, employ diverse assessment practices, these approaches are often tailored to their specific domains. This leads to a necessary divergence in criteria used to evaluate creativity, reflecting the distinct nature of creativity in each discipline. For instance, an engineering instructor might prioritize innovative problem-solving and technical feasibility, while an art instructor may focus on originality and aesthetic appeal. This variation in evaluation criteria, though appropriate for each domain, shows the importance of recognizing that creativity is not a one-size-fits-all concept. Baer (2012) notes that "validation studies of such supposedly domain-general tests have in fact lent support to the theory of domain specificity" (p. 22). This indicates that applying overly general criteria across disciplines can lead to misinterpretations of students' creativity, as such tests fail to capture the unique expressions of creativity relevant to specific fields, such as engineering or art. Consequently, pioneers in creativity research emphasize that creativity is inherently domain-specific, and assessments should be carefully designed to reflect the unique demands and contexts of each discipline (e.g., Kaufman, 2012). Therefore, the lack of clear, domain-appropriate standards—rather than general standards—makes it difficult to accurately assess creativity in its many forms, leading to an incomplete portrayal of students' capabilities and the effectiveness of instructional strategies.

## Literature Review

### *Creativity Assessments*

Creativity measurement generally falls into two categories: self-report measures and assessments by external judges (Kaufman et al., 2008). Self-reports are valued for capturing subjective experiences—such as motivations and beliefs (Silvia et al., 2012)—but often fail to account for domain differences (Kaufman et al., 2008). In contrast, external assessments offer standardized evaluation of creative products but require more effort and expertise (Li et al., 2024).

Standardized creativity measures include the Torrance Tests of Creative Thinking (TTCT), Alternative Uses Test (AUT), Remote Associates Test (RAT), Creative Product Semantic Scale (CPSS), Consensual Assessment Technique (CAT), and Wallach-Kogan Tests. These tools primarily assess general creativity through divergent or convergent thinking dimensions. Although labor-intensive, external assessments are regarded as the most accurate, as expert judgment remains central to evaluating creativity (Kaufman, 2016).

Although there is debate about the complexity and challenges of creativity assessments conducted by external evaluators due to the required human effort for scoring (Li et al., 2024), Kaufman et al. (2008) assert that "The best assessment of creativity in any field is usually the collective judgment of recognized experts in the field" (p. 52), highlighting the critical role that *assessment conducted by others* or expert evaluation plays an important role on assessing creativity. This view is supported by the example of any Nobel Prize awarding process, which relies on expert consensus to identify the most groundbreaking contributions across disciplines (Kaufman, 2016).

Assessments like TTCT, AUT, and Wallach-Kogan Creativity Tests are primarily designed to measure general creative potential relying on cognitive processes (i.e., dimensions of divergent thinking, associative thinking, imaginative expression etc.) rather than domain-specific expertise (Plucker & Makel, 2010). For example, Runco & Acar (2012)'s extensive review study looked at the quantitative (e.g., correlations between divergent thinking scores and creative achievements) and qualitative aspects (e.g., the ability to capture different dimensions of creativity) of these tests, concluding that while divergent thinking is a strong predictor of creative productivity, it does not capture the full spectrum of factors such as motivation, context, and specific scoring criteria. Consequently, these tests often fall short in engineering contexts, where creative problem solving is tightly coupled with technical knowledge, design constraints, and interdisciplinary integration. This limitation has led researchers to caution against relying solely on such general measures, advocating for a more holistic approach that includes cognitive, environmental, and motivational factors to better assess domain-specific creativity (Hocevar & Michael, 1979; Baer, 2011; Kim, 2006).

*Domain-Specificity of Creativity*

While creativity is often perceived in a broad, domain-general sense, careful examination and targeted questioning reveal that our intuitive understanding points toward a more domain-specific perspective (Baer, 2016). The hierarchical structure of the Amusement Park Theoretical (APT) Model of Creativity (Baer & Kaufman, 2005, 2017) presents us how creativity comes with levels/layers that integrates elements of both domain-general and domain-specific perspectives of creativity. The APT Model begins with *the foundational requirements for creativity*, encompassing domain-general factors like basic intelligence, motivation, and a supportive environment conducive to exploration and innovation. It then progresses to g*eneral thematic areas*, representing broad categories such as artistic or scientific creativity that share thematic commonalities. This is followed by *domains*, which refine these categories into specific disciplines, such as chemistry within the sciences or visual arts within the arts. Finally, *micro-domains* represent the most specific level, including specialized forms such as different types of poetry or subfields within psychology. This layered framework has directly informed the development of ECAT's dimensions in this study. For example, ECAT's measures of fluency and originality capture the creative output at the micro-domain level, reflecting the detailed, domain-specific nuances. Meanwhile, dimensions like cognitive flexibility and creative strengths (encompassing aspects such as usefulness, aesthetics, and colorfulness) align with the broader domain-general and thematic levels by assessing personal and environmental aspects. In this way, ECAT's multi-dimensional approach will mirror the hierarchical progression of creativity described by the APT Model, providing a structured and explicit linkage between theoretical understanding and practical assessment in engineering contexts.

Baer (2015) states that "different measures of creativity rooted in different domains will predict creative performance only in their respective domains" finds support in studies like Han & Marvin's (2002)'s work. They explored the domain-specific nature of creativity in young children across distinct creative areas, including storytelling, collage-making, and math word-problem creation, employing performance-based assessments and divergent thinking tests such as the Wallach-Kogan Creativity Test. Their results highlighted significant variability in creative performance among these domains, illustrating that children's creativity was highly context dependent. Notably, the use of general divergent thinking measures failed to predict performance consistently across the varied domains, further demonstrating the limitations of domain-general assessments and reinforcing the notion that creativity manifests differently based on the specific domain in question.

Similarly, Acar et al. (2024) revealed with their meta-analytic confirmatory factor analysis that creativity measures predict performance only within their domains through observed correlation patterns. Analyzing the Kaufman Domains of Creativity Scale (K-DOCS) of 45 correlation matrices from 30 studies, encompassing a sample size of 31,136 participants, they showed that the correlation between artistic and performance domains was the strongest (r = .478), while the correlation between every day and scientific/mathematical domains was the weakest (r = .178). This differentiation suggests limited overlap across distinct creative domains, affirming that measures within one domain do not reliably predict creative performance in unrelated domains, aligned with Baer's assertion.

Additionally, TTCT revealed that divergent thinking is often best conceptualized through a domain-specific lens, as performance across verbal and figural measures of TTCT can diverge significantly (Cramond et al., 2005). Statistically speaking, the divergence is evidenced by a low correlation coefficient between the scores on these two measures. Specifically, earlier findings reported by Cramond et al. (2005) highlighted a correlation of approximately r = 0.06 between verbal and figural test performances. Such a low correlation suggests minimal overlap in the creative abilities assessed by these two forms of the TTCT, indicating that they may be measuring different aspects of creativity rather than a cohesive, unified construct. This literature evidence supports the argument that meaningful creativity assessment must be tailored to specific contexts, presenting the complexity and specificity of creative capacities across different areas (Baer, 2015; Runco, 2004).

*Creativity Assessments in Engineering Education*

To address creativity in education, one must be able to measure it. We review existing approaches to creativity assessment in engineering and their limitations. Over the past few decades, a growing body of research has focused on assessing creativity within engineering education (e.g., Charyton et al., 2008; Denson et al., 2015; Kershaw et al., 2019; Titova & Sosnytska, 2020). Studies in this area have explored various approaches to creativity assessment, including self-reports (e.g., Atwood & Pretz, 2016; Charyton & Merrill, 2009; Deo et al., 2019), peer evaluations (e.g., Telenko et al., 2015), instructor assessments (e.g., Tekmen-Araci & Mann, 2018), and product-based assessments (e.g., Charyton et al., 2011). Each method contributes unique insights into understanding and measuring creativity within the engineering domain. However, a heavy reliance on certain measures, such as self-reports, can present notable disadvantages. For example, Pretz & McCollum (2014) argue that elf-reports are susceptible to biases such as social desirability, where students may overestimate their creative abilities to conform to perceived expectations. Furthermore, self-report measures may fail to capture the observable outcomes of creativity, particularly in task-based environments like engineering, where the optimal level of creativity is found in tangible tasks, products, designs. For instance, Kazerounian and Foley (2007) used self-reports to explore students' perceptions of creativity in engineering as it relates to technical skills and adaptability. They found that students often struggled to identify creativity in their engineering work, perceiving it as secondary to technical competencies. Their study highlighted students' biases toward technical performance over creative approaches, suggesting that self-reports might reveal attitudes that undervalue creativity in engineering contexts.

Studies on assessing engineering creativity typically focus on either the creative individual or the creative product. While the creative process is a more complex phenomenon, where divergent thinking plays a crucial role in assessment, a comprehensive process assessment is even more challenging. It requires evaluating problem definition and selection, both of which significantly influence the quality of the final solution.

Since engineering is largely centered on design, the product often takes precedence. The product serves as a unifying factor between engineering and creativity; thus, research highlights that engineering educators may be primarily interested in assessing the level of creativity exhibited in final products (Lerdal et al., 2019). However, evaluations of both the creative process and product are not based on self-assessments or simple questionnaires. Instead, they require substantial time for completion and scoring, along with the expertise necessary for accurate evaluation. Consequently, the resulting scores provide deeper insights into participants' true creative abilities.

In addition to the well-established TTCT, another instrument that has set the standard for creativity assessment is the Consensual Assessment Technique (CAT). Specifically in engineering, a variation of the CAT, known as the Creative Solution Diagnostic Scale (CSDS), was developed to assess creativity in engineering design products by focusing on four key dimensions: effectiveness, novelty, elegance, and genesis (Cropley & Kaufman, 2012). Similarly, Denson et al. (2015) used a web-based CAT as a reliable and valid assessment method to evaluate engineering student products in terms of three major dimensions: creativity, technical strength, and aesthetic appeal. Another widely recognized tool is the Creative Engineering Design Assessment (CEDA), which also evaluated creativity in engineering design (Charyton et al., 2008). Unlike CSDS and web-based CAT, which emphasized the final product, CEDA focused on the creative process by assessing divergent and convergent thinking, placing particular emphasis on an individual's creative abilities throughout the creative process.

The evolution of creativity assessment in engineering education has demonstrated the strengths and limitations of various existing tools. While product-based assessments, such as the CSDS and web-based CAT offer structured evaluation methods, they primarily focus on the final design outcome, often overlooking the cognitive processes that lead to creative solutions. Conversely, tools like the CEDA emphasize the creative process, assessing both divergent and convergent thinking, but lack the structured evaluation frameworks and broad accessibility of web-based CAT adaptations. This gap suggests a critical need for an alternative engineering creativity assessment tool that integrates the strengths of these existing approaches. The alternative assessment tool can be web-based for accessibility and scalability, while also being rooted in the reliability of the CAT methodology and capable of evaluating engineering creativity holistically by incorporating both product and process dimensions. Such an instrument would provide a more holistic understanding of creativity in engineering, capturing not only the novelty and functionality of final products but also the iterative thought processes, problem formulation strategies, and cognitive flexibility. These are essential yet often neglected components that can truly define a unique engineering product.

*Consensual Assessment Technique (CAT)*

Unlike rigid rubrics or predefined scoring systems, the CAT relies on the judgment of domain-specific experts who assess artifacts such as products, ideas, or solutions. In engineering, creativity often manifests through divergent thinking, innovative combinations of ideas, and the contextual appropriateness of solutions. These qualities are challenging to quantify but are crucial for solving complex engineering problems. CAT addresses this challenge by leveraging expert evaluations to ensure that assessments remain relevant to the specific domain. Unlike objective tools, which are adept at measuring technical correctness or efficiency, CAT excels in capturing the nuanced and multi-faceted aspects of creativity assessment.

While CAT is a subjective assessment method, it incorporates mechanisms to ensure reliability and minimize bias, making it both valid and robust. A key feature of CAT is the use of multiple independent evaluators who assess artifacts without external influence or collaboration. Their judgments are aggregated to produce a consensus-based evaluation, reducing the impact of individual biases. Additionally, studies utilizing CAT frequently measure inter-rater reliability, validating the consistency of expert judgments and reinforcing the credibility of the assessments (Barth & Stadtmann, 2021).

In engineering education, CAT has been applied to assess the novelty and functionality of prototypes in settings such as capstone projects and design challenges (Buelin-Biescker & Denson, 2013; Denson et al., 2015). For example, while Denson et al.'s instrument assessed creativity in design projects, it grouped creativity into a single score alongside technical and aesthetic scores, thus, not differentiating the contributors to creative performance. Charyton's CEDA, on the other hand, incorporated divergent thinking tasks but was adapted from general creativity tests and did not use expert judgment of student-generated products. These approaches either lack dimensional detail or lack context specificity, which ECAT aims to provide by combining a CAT approach with structured divergent-thinking criteria. Additionally, the concept of creative productivity in college-level engineering education had not been explored through an adapted CAT before. Therefore, this study develops the Engineering Creativity Assessment Tool (ECAT) to address the unique challenges of creativity evaluation in this field. Unlike objective tools, which may oversimplify or overlook the complex interplay between imagination and functionality, ECAT aims for a comprehensive evaluation that encompasses both. By combining expert judgment with engineering-relevant criteria, ECAT seeks to build on the strengths of CAT, ensuring that assessments reflect both creative thought processes and the practical implementation of innovative ideas.

## Proposed Framework for ECAT

Given the importance of creativity in engineering, this study leveraged the flexibility of the original CAT and tailored it for use in an engineering context. Historically, the focus on creativity in engineering has been linked to the invention of the concept of divergent thinking (Guilford, 1967). However, the notion of attributing innovative problem-solving in engineering design scenarios solely to a single cognitive ability (divergent thinking) has recently been conceptually challenged (Acar & Runco, 2019; Baer, 2012, 2016, 2017) and empirically (Carpenter, 2016; Kim, 2011). Existing methods either rely on students' own perceptions or evaluate only end products, missing the nuanced creative processes in engineering design. Thus, a tailored approach like ECAT is warranted to capture those nuances.

Rather than presuming a universal creative ability with divergent thinking as its core component, domain-specific perspectives on creativity propose that a diverse range of skills relevant to a particular domain or task are essential for achieving creative outcomes within that context (Baer, 2012, 2016). This perspective acknowledges the potential role of divergent thinking in creative achievement but emphasizes the importance of understanding its contribution within the specific context of a domain. Instead of treating divergent thinking scores as definitive indicators of an individual's creativity, domain-specific perspectives suggest that the extent to which divergent thinking contributes to creative behavior in a given domain should be determined through empirical investigation (Denson et al., 2015). This rationale motivates the development of ECAT, which evaluates the domain-specific nature of engineering products while maintaining the validity and reliability of CAT-based creativity assessments. In other words, the level of creativity of an engineering product depends on domain-familiar judges.

According to the systems concept of functional creativity (Cropley & Cropley, 2010), successful engineering creativity must integrate technical feasibility, functionality, and innovation. ECAT also aims to reflect this view by incorporating the evaluation of not only technical and innovative aspects of creative performance but also general indicators of creativity such as flexibility, elaboration, tolerance to ambiguity, and other categories of creative strengths that can be evaluated in the context of engineering. These are dimensions often overlooked in traditional engineering assessments, which tend to prioritize final products rather than the creative thinking and problem-solving strategies involved.

## Methods

ECAT was developed to evaluate the creative productivity of engineering student products. Prior to ECAT scoring, students (n = 199) completed a demographic questionnaire including age, sex, and college major; race/ethnicity was not collected. Participants were undergraduates at a northeastern U.S. public university, with 56.3% male and 43.7% female, ranging in age from 18 to 33.

The sample included students from various majors: Biomedical Engineering (17.6%), Civil (12.6%), Chemical (7.5%), Computer Science & Engineering (12.6%), Electrical (5.0%), Environmental (4.0%), Manufacturing & Engineering Management (7.5%), Materials Science (5.5%), Mechanical (24.1%), Engineering Physics (0.5%), Dual Major (1.0%), Undeclared (1.5%), and Unknown (0.5%).

**Instrument**

*ECAT Dimensions*

ECAT builds on the versatile CAT adapting it for engineering education by integrating domain-specific criteria and divergent thinking components. While prior CAT-based studies clustered creativity into creativity, technical strength, and aesthetic appeal (Denson et al., 2015), ECAT expands this framework to include four key dimensions: fluency, originality, cognitive flexibility, and creative strengths.

**Fluency.** It measures the quantity and variety of ideas in a product, focusing on diversity of components, functional features, and variations in design. Unlike traditional fluency measures, ECAT evaluates both idea generation and feasibility. Subcategories include:

1. Material diversity – variety and number of materials used,
2. Functionality – presence of distinct functional elements,
3. Variations – range of unique design strategies.

High fluency reflects both divergent thinking and coherent integration of components.

**Originality.** It evaluates how novel or unconventional a solution is, especially in relation to common engineering practices. Subcategories:

1. Uniqueness – divergence from standard solutions,
2. Innovation – new functionalities or enhancements,
3. Surprise – unexpected use of materials or configurations.

A high originality score reflects radical innovation and departure from norms.

**Cognitive Flexibility.** ECAT's cognitive flexibility combines flexibility**,** elaboration, and resistance to premature closure (RPC) to capture adaptability and refinement in problem-solving (see Reiter-Palmon et al., 2019). It includes:

- **Flexibility:** Assesses adaptability to constraints. High scores reflect coherent, integrated design shifts, while low scores suggest rigidity.
- Elaboration: Measures the depth and completeness of the design. Subcategories:
    1. Functional completeness – how well the product meets its goals,
    2. Depth of explanation – clarity of rationale and design trade-offs.
- **RPC:** Captures persistence in refining ideas. Subcategories:
    1. Depth of development – improvements across iterations,
    2. Approach to uncertainty – willingness to explore beyond the obvious.

High cognitive flexibility reflects iterative thinking and tolerance for ambiguity. All subdimensions are scored on 5-point or descriptive scales.

**Creative Strengths (CS).** Adapted from TTCT's original framework, ECAT's creative strengths evaluate the expressive and practical impact of the design. Subcategories:

1. Usefulness – relevance and applicability,
2. Aesthetic appeal – structural and visual design quality,
3. Colorfulness – effective use of color to enhance function or clarity.

These were assessed via images/videos using descriptive scales. High scores signal well-executed, impactful designs.

Figure 1 and Table 1 provide a worked example using a hypothetical smart urban lighting system, demonstrating how each ECAT dimension can be applied to evaluate creative engineering output.

**Figure 1**

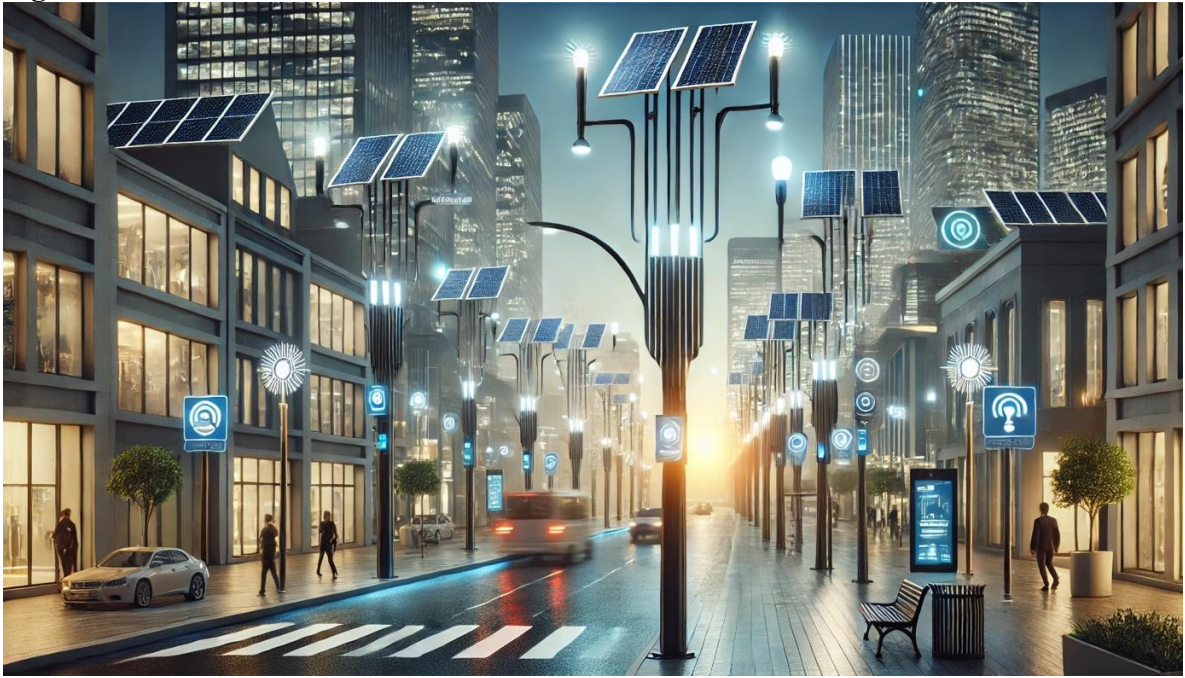

Figure 1 shows the illustrated summary of the hypothetical smart urban street lighting system generated by DALL·E (Open AI, 2025). This hypothetical engineering solution is solar-powered and sensor-integrated. It adapts lighting based on pedestrian and vehicle activity, collects environmental data, and enhances urban aesthetics through dynamic color effects. Table 1 shows how ECAT dimensions are used to score this highly creative engineering solution for urban life settings.

**Table 1**

*ECAT Scoring Table*

| Dimension | Subdimension | Score | Justification |
|---|---|---|---|
| Fluency | Material Diversity | 5 | Uses a wide range of materials: recycled aluminum, solar panels, sensors, polymers, and LED components. |
| | Functionality | 5 | Integrates multiple functional components: lighting, sensors, data collection, and wireless communication. |
| | Variations | 5 | Offers modular design variations tailored for different urban zones (e.g., parks, main roads). |
| Originality | Uniqueness | 5 | Diverges from traditional street lighting by combining sustainability, smart tech, and data collection. |
| | Innovation | 5 | Introduces new functionalities like dynamic lighting response, real-time analytics, and urban customization. |
| | Surprise Factor | 5 | Unexpected integration of environmental sensing and community-responsive features in a conventional infrastructure product. |
| Cognitive Flexibility | Flexibility | 5 | Adapted the design based on evolving constraints (urban space, lighting regulations, energy availability). |
| | Elaboration – Functional Completeness | 5 | Thorough technical documentation, fully integrated system with energy, sensor, and communication modules. |
| | Elaboration – Depth of Explanation | 5 | Detailed rationale behind design decisions, including trade-offs between durability, cost, and energy efficiency. |
| | RPC – Depth of Development | 5 | Evolved through multiple prototypes and feedback cycles; significantly enhanced over initial design. |
| | RPC – Approach to Uncertainty | 5 | Embraced novel ideas like real-time data use and dynamic color lighting, rather than relying on conventional LED systems. |
| Creative Strengths | Usefulness | 5 | Practical for city use: improves safety, lowers energy use, and adds urban functionality through data analytics. |
| | Aesthetic Appeal | 5 | Sleek and modern design that enhances the visual quality of urban infrastructure. |
| | Colorfulness | 5 | Dynamic use of color in lighting to indicate events, emergencies, or enhance public engagement. |

**Procedures**

*Selection of Evaluators*

Since a reliable judgment of domain-specific creativity requires an appropriate group of evaluators, we recruited five quasi-experts in the engineering domain to assess the creative productivity of student engineering products. Our team included four senior-level graduates level engineering students. In alignment with Amabile's (1996) suggestion, we ensured that the evaluators' level of expertise in engineering was significantly higher than that of the students who created the products. Also, relying on previous evidence (Kaufman et al., 2013), we decided that quasi-experts were suitable evaluators for this study since their expertise level and the context of the product that they evaluate importantly intersect.

All evaluators were available and committed to all stages of the assessment process, including initial training on the ECAT, calibration meetings to ensure consistency, actual scoring, and follow-up meetings. This commitment to the full assessment process and consistency was critical for reliable scoring. Additionally, the diversity in their educational levels and the inclusion of both academic and collaborative perspectives aimed to mitigate individual biases and enrich the evaluation.

*Calibration Meetings*

To establish the reliability and validity of ECAT, we trained our quasi-expert evaluators to calibrate their scoring practices and conducted pilot tests to refine the assessment process. This stage included three calibration meetings.

The first meeting introduced evaluators about the concept of creativity and its evolving role throughout the history of engineering education. This meeting familiarized evaluators with ECAT's purpose and dimensions, addressing each dimension separately to discuss its relevance, supported by examples of engineering products. Additionally, evaluators participated in a hands-on workshop where they created simple engineering products using the same materials available to students. This activity fostered empathy for the challenges students face and provided evaluators with practical experience to internalize the ECAT dimensions.

The second calibration meeting focused on refining ECAT based on evaluators' feedback. This meeting basically assessed whether evaluators had any fundamental misunderstandings regarding their ECAT administration. A week before the meeting, evaluators received three samples of student products that demonstrated low, medium, and high creativity. They were asked to review and evaluate these products within a week using a Qualtrics form. After collecting the evaluations, we shared their score distribution. We highlighted significant deviations and facilitated a group discussion, encouraging each evaluator to explain their reasoning behind assigned scores. Anecdotal notes from this session helped clarify any misunderstandings about the scoring criteria, ensuring evaluators were aligned in their interpretation by the end of the meeting. This session aimed to resolve rubric ambiguities based on evaluators' insights and data, which were then applied to further refine ECAT in preparation for the third meeting.

The final calibration meeting assessed whether evaluators had developed sufficient consistency to capture the creative productivity of students' engineering products, shifting the focus from comprehension of ECAT dimensions to reliable application. We provided five additional samples with both low and high creative productivity and expected evaluators' scores to align closely. The meeting concluded with evaluators largely consistent in their judgments, with only minor disagreements such as unclear rubric language, borderline cases in scoring, personal preferences in highly subjective items like aesthetics.

*Evaluation of Engineering Products*

The ECAT was employed to assess the creative productivity of student generated products developed between Fall 2018 and Spring 2019. Each evaluator was assigned an individual rater ID to ensure anonymity and consistency throughout the evaluation process. Evaluators accessed the ECAT via a Qualtrics link. The evaluation period commenced in November 2025 and spanned three months, providing raters with autonomy in pacing while adhering to the deadline set by the research team.

**Data Analysis**

Before conducting formal data analyses, we had three sessions of calibration meetings to ensure consistency and alignment among the raters when implementing ECAT. Calibration sessions allowed raters to review the assessment criteria, discuss potential ambiguities, and establish a shared understanding of scoring rubrics. Given the subjective nature of ECAT, achieving absolute agreement was not the primary goal; rather, ensuring a consistent interpretative framework among raters was the key objective. The results from the calibration session confirm that raters reached a sufficient level of alignment, allowing us to proceed confidently with the actual evaluation process. A high proportion of partial agreement (up to 71.4%) suggested that raters had a shared understanding of the scoring criteria, even if they did not always assign identical scores. The level of disagreement was within an acceptable range (within 28.6% – 35.3%), confirming that ECAT rubric was interpretable but still flexible enough to accommodate diverse expert perspectives.

*Reliability*

To ensure consistent evaluator interpretation, we conducted calibration meetings where experts aligned their scoring strategies for creative engineering products. Given the subjective nature of creativity assessment, achieving absolute agreement was not expected; instead, the goal was to ensure consistency in scoring interpretations.

During calibration sessions, inter-rater agreement reached 71.4%, allowing us to proceed with the full evaluation phase. To quantify inter-rater reliability, we computed Kendall's Tau-b correlations between sum scores of each subscale. Correlations between 0.40 and 0.70 were considered indicative of moderate to strong agreement, and our results fell within this range ($p < 0.01$), supporting consistent evaluator judgment across ECAT dimensions. This aligns with prior research, which suggests that agreement among independent raters reinforces construct validity in creativity assessment (Denson et al., 2015).

After the evaluation phase, we assessed internal consistency by computing Cronbach's Alpha ($\alpha$) across ECAT subscales. Using the commonly accepted benchmark that $\alpha \geq 0.70$ indicates acceptable reliability (Nunnally & Bernstein, 1994), our result ($\alpha = .89$) demonstrated strong internal consistency, suggesting that ECAT subscales reliably measure their intended constructs.

*Validity*

**Content Validity.** To ensure content validity, we conducted structured calibration sessions with expert evaluators. The process began with a systematic item review, in which experts assessed whether ECAT's six subscales comprehensively captured key aspects of engineering creativity. Rather than using individual content validity index (CVI) scores, we employed group discussions to elicit qualitative feedback, which was documented and systematically analyzed. Based on expert recommendations, certain items were refined or removed to enhance ECAT's clarity and construct representation.

**Convergent Validity.** To evaluate convergent validity, we compared ECAT scores with TTCT scores from the same sample of students. We assessed whether ECAT dimensions correlate with TTCT's established measures of divergent thinking, including verbal fluency, verbal originality, and figural elaboration.

The results provided partial support for convergent validity, demonstrating that ECAT's originality ($r = 0.26–0.32$), elaboration ($r = 0.27$), and flexibility ($r = 0.29$) moderately correlated with TTCT's verbal creativity measures. These findings suggest that ECAT captures similar aspects of creative productivity as TTCT verbal subscales.

As expected, correlations between ECAT and TTCT figural measures were lower ($r < 0.20$), reinforcing that engineering creativity may involve domain-specific manifestations distinct from traditional TTCT assessments. This is because ECAT is designed to assess creativity through the lens of engineering design whereas TTCT figural tasks focus on visual-spatial abilities such as fluency, elaboration, and originality in abstract drawing tasks. These differing emphases lead to the expectation that performance on TTCT figural tests would not strongly correlate with engineering-focused creativity measures. This aligns with prior research indicating that engineering creativity assessments emphasize problem-solving and functional originality, whereas TTCT figural tasks assess visual fluency and elaboration (Reiter-Palmon et al., 2019).

**Discriminant Validity.** To test discriminant validity, we examined Pearson correlations among ECAT subscales to determine whether each dimension captures a distinct aspect of engineering creativity.

The results confirmed that fluency, originality, and flexibility are related but remain statistically distinct constructs, as indicated by moderate correlations between fluency and originality (F2–O1: $r = 0.52$, F3–O2: $r = 0.72$) and weaker correlations between fluency and flexibility (F1–FL2: $r = 0.22$). This distinction supports the theoretical framework that fluency represents idea quantity, originality reflects uniqueness, and flexibility captures adaptability in engineering design (Reiter-Palmon et al., 2019).

Furthermore, elaboration was found to be independent from fluency and originality, reinforcing that depth and refinement of ideas do not necessarily align with idea generation or novelty. This is evidenced by lower correlations between elaboration and fluency (F1–E1: $r = 0.26$, F2–E1: $r = 0.47$) and moderate correlations with originality (O1–E1: $r = 0.42$, O2–E2: $r = 0.45$).

Additionally, resistance to premature closure (RPC) emerged as a separate construct, exhibiting weaker associations with fluency (F1–RPC1: $r = 0.30$, F2–RPC1: $r = 0.39$) and flexibility (FL1–RPC1: $r = 0.45$). Lastly, creative strengths (CS) remained statistically distinct from fluency and flexibility, with aesthetic appeal (CS4) showing stronger associations with originality (O1–CS4: $r = 0.44$, O2–CS4: $r = 0.45$).

It is important to note that we averaged the raters' scores on each item for each student's product to obtain a single ECAT score per dimension. These findings confirm that ECAT's dimensions measure theoretically distinct constructs.

## Results

**Exploratory Factor Analysis (EFA)**

To determine the underlying structure of the dataset and identify the factor loadings of each subscale in the ECAT, we conducted an Exploratory Factor Analysis (EFA) using Python. As a prerequisite for factor analysis, we first performed Bartlett's test for sphericity, which was significant, $\chi^2(197) = 537.02$, $p < .001$. This result indicated that the correlation matrix significantly differed from an identity matrix, confirming the suitability of the dataset for factor analysis.

To identify the appropriate number of factors to retain, we used multiple criteria, including eigenvalues from Principal Component Analysis (PCA), the scree plot method (Cattell, 1966), and theoretical expectations. The initial PCA revealed that the first four factors collectively accounted for 74.31% of the total variance. While the Kaiser criterion suggested a two-factor solution, we chose a four-factor model based on the scree plot's visible "elbow" at the fourth component and theoretical considerations. Prior research cautions that the Kaiser criterion may under-extract factors, especially when theory supports a more nuanced construct structure (Zwick & Velicer, 1986). Figure 2 illustrates this elbow, indicating diminishing explanatory value beyond the fourth factor.

**Figure 2**

*Scree plot of principal components*

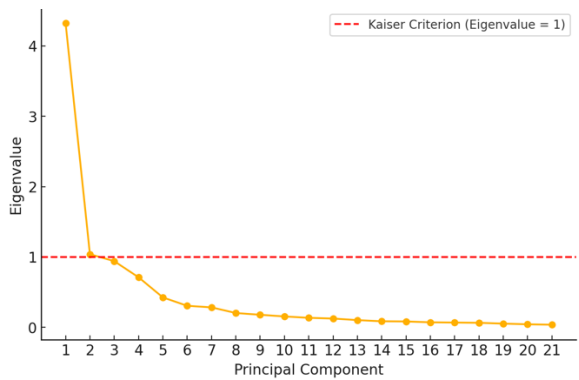

A subsequent EFA with a four-factor solution is visualized in Figure 3 through a heatmap of item loadings. Following established behavioral and social science guidelines, factor loadings $p \geq 0.40$ were considered substantive (Boateng et al., 2018). The color intensity and numerical values within each cell represent the strength and sign of the item loadings on each factor, respectively.

**Figure 3**
*Heatmap of factor analysis results*

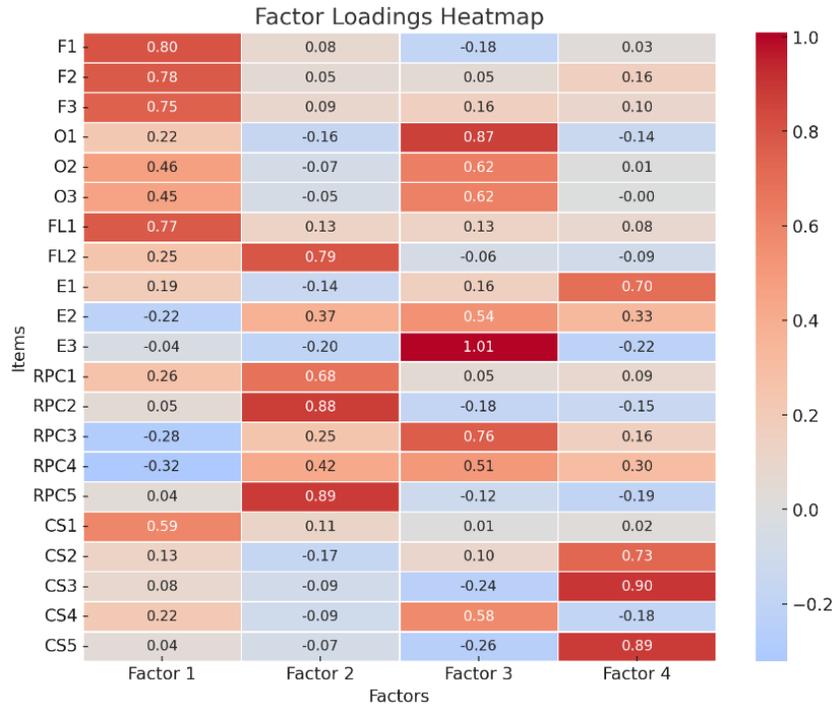

The results indicate a coherent thematic grouping as follows:

Factor 1: Fluency
This factor aligns with the ability to generate multiple distinct ideas or products, a core dimension in both creativity theory and prior assessments (e.g., Guilford, 1967). Its emergence as a distinct factor confirms the theoretical expectation that idea generation (fluency) is a foundational but separable component of creative output.

Factor 2: Cognitive Flexibility
This factor constitutes Flexibility items and RPC items, especially FL3, FL4, and RPC2. combines flexibility, the capacity to shift between different ideas or perspectives, with resistance to closure, which reflects tolerance for ambiguity and openness to multiple possibilities. The co-loading of these items suggests a shared cognitive mechanism related to adaptive thinking under constraints. The relabeling of this factor as "Cognitive Flexibility" is supported by creativity literature linking flexibility and resistance to premature closure as manifestations of divergent and non-linear thinking (Runco & Acar, 2012).

Factor 3: Originality
The main loadings under this factor are Originality items, E3, and RPC3 (to a lesser extent). This factor represents the novelty and uniqueness of ideas. Early theoretical models emphasize that originality often co-occurs with unusual combinations or interpretations (Sternberg, 2003), which can explain the cross-loading of elaboration here.

Factor 4: Creative Strengths
This factor includes multiple CS along with an elaboration item. It captures the extent to which ideas are developed with detail, refinement, and expressive strength. The association between elaboration and creative strengths (e.g., usefulness, aesthetics) reflects the developmental depth of ideas, consistent with new study findings that highlight elaboration as a vehicle for idea execution and communication (e.g., Gillier et al., 2024).

The observed overlap between RPC and FL2 suggests a conceptual linkage between RPC and adaptation to constraints (FL2). Consequently, we redefined Factor 2 as *Cognitive Flexibility*, encapsulating both flexibility and RPC dimensions. This redefinition is theoretically grounded: both constructs reflect an individual's ability to tolerate ambiguity, shift perspectives, and explore multiple possibilities before arriving at a solution (Well, 2024). While flexibility refers to generating diverse ideas or categories, RPC involves resisting early closure and remaining open to alternative interpretations. Their combination reflects a shared underlying cognitive disposition toward adaptive, divergent thinking, reinforcing the validity of grouping them together beyond statistical justification. Additionally, the distinction between Factors 1 and 3 reinforces that fluency and originality in ECAT represent separate constructs. To confirm this factor structure, a Confirmatory Factor Analysis (CFA) was subsequently conducted.

**Confirmatory Factor Analysis (CFA)**

The initial CFA model was evaluated using key fit indices to assess its alignment with the hypothesized four-factor structure. The model demonstrated an acceptable yet suboptimal fit, as indicated by a Comparative Fit Index (CFI) of 0.88, a Normed Fit Index (NFI) of 0.86, an Adjusted Goodness-of-Fit Index (AGFI) of 0.80, and a Tucker-Lewis Index (TLI) of 0.84. Additionally, the Root Mean Square Error of Approximation (RMSEA) was 0.127, and the model's chi-square value was $\chi^2 = 199.47$, $p < 0.01$, suggesting room for improvement.

To enhance the model's fit, Factor 2 was refined by integrating flexibility and RPC into a single construct, while E3 was removed from Factor 3 to mitigate its inflating effect on the error terms. These modifications led to a substantial improvement in model fit, with the CFI increasing to 0.95, NFI to 0.93, AGFI to 0.90, and TLI to 0.93. The RMSEA was reduced to 0.089, and the chi-square value improved to $\chi^2 = 97.42$, $p < 0.01$, confirming a better alignment with the data. E3 showed a complex pattern, loading with creative strength items in EFA. We ultimately aligned it with the cognitive flexibility construct in the CFA, reasoning that elaboration (adding detail) conceptually fits with an adaptive, iterative problem-solving mindset, whereas the creative strengths factor was intended to capture output qualities like aesthetics and usefulness. The improved CFA results reinforced the validity of the revised four-factor model, strengthening the theoretical foundation of the ECAT structure.

It is also important to note that, since another sample was not available, the CFA was conducted on the same dataset, so results, while promising, should be confirmed with new data in future research. We achieved an excellent fit (CFI 0.95, etc.). This result may be somewhat optimistic because the model was tweaked based on the idiosyncrasies of this sample. However, we strongly suggest future researchers cross-validate the factor structure on a hold-out sample or in a new study to ensure its stability. Additionally, our decision to combine flexibility and RPC into one factor and to drop one elaboration item (E3) in the CFA model refinement should be justified in terms of content. Our suggestions for the modifications were theoretically sound, not just statistically driven because RPC has overlapping concepts with flexibility, and E3 has more to do with the uniqueness of the engineering problem, which was highly overlapping with the originality dimension of the instrument. Originality already captures the uniqueness aspect of creativity therefore elimination of E3 was needed.

## Discussion

In this study, we try to provide support for the reliability and validity of ECAT in evaluating creative productivity of engineering students. The factor structure identified through EFA and confirmed via CFA reinforced the distinct but interrelated nature of fluency, originality, cognitive flexibility, and creative strengths aspects of engineering creativity. Our redefined factor structure aligns with prior research highlights. For instance, Charyton et al. (2018) emphasized the role of divergent thinking, fluency, originality, and flexibility in engineering creativity. Similar to ECAT their instrument CEDA captures multiple dimensions of creativity; however, CEDA primarily focuses on psychometric approaches adapted from traditional creativity assessments like TTCT. Notably, our results suggested that ECAT's cognitive flexibility dimension, which includes flexibility, elaboration and resistance to RPC is a novel way of combining these three divergent thinking components that aligned in the context of engineering product design.

The ECAT framework also shares conceptual similarities with Denson et al. (2015), who developed an instrument for assessing creativity in engineering design. Denson et al. focused on three primary dimensions: creativity, technical strength, and aesthetic appeal, which parallels ECAT's structure. However, a key distinction lies in the level of granularity. More specifically, ECAT decomposes creativity into fluency, originality, and cognitive flexibility, creative strengths each with multiple subcomponents, thereby offering a more detailed analysis of engineering creativity. Denson et al. (2015)'s finding supports the notion that while general creativity tests capture some aspects of engineering creativity, there are significant domain-specific elements (particularly those captured by ECAT's cognitive flexibility and elaboration metrics) that require specialized assessment. Also, our findings reinforce domain-specific creativity theory: performance on ECAT did not strongly correlate with general (figural) TTCT scores, aligning with Baer's (2015) assertion that creativity doesn't fully generalize across domains.

Additionally, Denson et al. (2015) emphasized expert raters' reliability in creativity assessments, a concern also addressed in this study. Our results indicate moderate to strong inter-rater reliability, supported by Kendall's Tau-b correlations and Cronbach's alpha of 0.89, reinforcing the dependability of quasi-expert evaluations. This is consistent with Denson et al.'s findings, which highlight that while absolute agreement is rare in creativity assessments, structured calibration meetings and systematic evaluation protocols improve scoring consistency.

Our refined CFA model demonstrated substantial improvements in model fit following the integration of flexibility and RPC into a single construct and the removal of E3 from Factor 3. These findings suggest that adaptability in design, as captured by flexibility and RPC, is conceptually linked, reinforcing the role of iterative refinement in engineering creativity. It is because iterative design inherently requires engineering designers to remain open to multiple possibilities and to adapt their solutions based on continuous feedback. The refined model aligns with prior findings that engineering creativity is not solely about ideation but also about persistence, adaptability, and the ability to explore alternative solutions before reaching closure (e.g., Cropley, 2015b).

## Limitations

Our study sample consisted of 199 undergraduate engineering students from a single northeastern public U.S. university, which may limit the generalizability of the findings. The sample lacked representation from other educational institutions, including private universities, international institutions, and community colleges. Additionally, while the study included students from diverse engineering disciplines, certain fields were underrepresented. Future research should expand the sample size and include students from a broader range of institutions and disciplines to assess ECAT's applicability across different engineering education contexts.

Although demographic data such as age, sex, and college major were collected, race/ethnicity information was not recorded. As previous research suggests that cultural and demographic factors influence creativity (e.g., Alizamar et al., 2019), the absence of this data limits the ability to examine potential differences in creative performance among diverse student populations. Future studies should incorporate a more comprehensive demographic profile to explore whether ECAT results vary across different racial, ethnic, or socioeconomic backgrounds.

Participants were given only 20 minutes to create their engineering products, which may not have provided enough time for them to fully develop, iterate, or refine their designs. In contrast, Charyton et al. (2008) emphasized the importance of longer creative problem-solving tasks to capture the depth of ideation and iterative refinement. Future research should explore longer task durations or multiple stages of evaluation to better capture creativity in engineering design.

Although quasi-expert evaluators were trained through calibration meetings to reduce subjectivity, achieving absolute agreement in creativity assessment remains challenging. Inter-rater agreement reached 71.4%, which, while acceptable, still suggests variability in scoring interpretations. Incorporating machine learning or AI-assisted evaluation tools to complement human judgment may improve objectivity and reliability in future ECAT implementations.

The ECAT assessment primarily focused on physical prototypes and their corresponding verbal explanations. However, in modern engineering practice, creativity often involves digital modeling, computational simulations, and software-based solutions (e.g., Robertson & Radcliffe, 2008), none of which were evaluated in this study. This limitation suggests that ECAT, in its current form, may not fully capture creativity in engineering fields that emphasize virtual prototyping, computational modeling, or interdisciplinary problem-solving. Future research should investigate how ECAT could be adapted for digital engineering creativity assessments.

The study only assessed creativity at a single point in time, preventing an analysis of how students' creative abilities evolve over the course of their engineering education. Prior studies (e.g., Cropley, 2015) emphasize that creativity in engineering is a dynamic process that develops over time, influenced by coursework, project-based learning, and professional experiences. Now that ECAT is available, researchers and instructors can design studies to see if certain teaching methods actually improve students' ECAT scores over time. Future research should implement longitudinal studies to track students' creative growth across multiple semesters or academic years, providing a more comprehensive understanding of how engineering education impacts creativity development.

## Implications

This study provides essential insights into the robust evaluation of engineering creativity, addressing the previously noted shortcomings in existing assessment methods. The development and validation of ECAT carry important implications for both pedagogical practices and educational research in engineering contexts. First, ECAT offers educators a reliable and nuanced framework for diagnosing creative competencies within their students. Implementing ECAT in a course would require training faculty or TAs to use the rubric reliably. Our study's calibration approach provides a model for how to achieve this (with 71.4% agreement in our case).

A realistic application of ECAT would be in a first-year engineering design course, where students are introduced to open-ended problem solving through hands-on team projects. In such courses, students often create physical prototypes using basic materials (e.g., cardboard, sensors, motors) to address user-centered design challenges. By using ECAT to evaluate student products, instructors can go beyond functionality to assess creative dimensions such as fluency, elaboration, originality, and cognitive flexibility. For example, if students in a section consistently score low on cognitive flexibility or RPC, instructors can introduce design prompts that require divergent thinking or encourage more iterative prototyping. This allows creativity instruction to be tailored based on real-time diagnostic feedback, rather than relying solely on final product performance or informal impressions. Additionally, educators can identify areas where students commonly struggle (such as originality, cognitive flexibility, or resistance to premature closure) and subsequently tailor instructional interventions to target these specific skill areas. Such targeted instructional practices can better prepare students for the dynamic demands of real-world engineering tasks, where creative problem-solving and adaptability are paramount.

Moreover, ECAT's novel conceptual integration (the merging of flexibility, elaboration, and resistance to premature closure) under the broader dimension of cognitive flexibility suggests potential refinements to existing creativity theories. For instance, Hirshfield & Koretsky (2020) found that engaging engineering students in a collaborative virtual environment cultivated creative thinking skills. This aligns with the idea that the process (flexibility, adaptation in a team setting) is key. Also, ECAT's integrated approach provides empirical support for theoretical perspectives asserting that engineering creativity significantly relies on adaptive problem-solving and iterative refinement, beyond merely generating novel ideas. Thus, ECAT not only serves as an assessment tool but also has a potential to advance our theoretical understanding of the construct of creativity within engineering, emphasizing the importance of iterative, adaptive cognitive processes alongside traditional divergent thinking.

Additionally, the detailed and validated structure of ECAT can serve as a benchmarking tool for curriculum development and instructional effectiveness in engineering education. Educators and curriculum designers can utilize ECAT assessments longitudinally to monitor how different teaching methodologies influence the development of students' creative abilities over time. This can lead to data-driven pedagogical reforms that more effectively nurture innovation skills critical for future engineering professionals. Moreover, educational institutions may employ ECAT in program evaluation to ensure alignment with industry demands, accrediting bodies' creativity expectations, and broader educational objectives centered on innovation and adaptability.

Furthermore, ECAT's comprehensive and domain-specific approach aligns well with current educational movements advocating for integrated STEM curricula and experiential learning environments such as makerspaces. ECAT's focus on both product and process makes it especially relevant for evaluating student work produced in project-based learning scenarios. Thus, ECAT has the potential to become a standard assessment tool in contexts that emphasize holistic, real-world engineering projects.

Lastly, this research highlights the critical role domain familiarity plays in accurate creativity assessment, suggesting implications for evaluator selection and training. Engineering educators or evaluators should be adequately trained and calibrated using standardized frameworks like ECAT, ensuring consistency and reliability in judgments of creativity. This reinforces the importance of domain-specific assessment tools and expert evaluators in effectively measuring complex constructs such as creativity.

# Conclusion

This study introduces and validates ECAT, a novel and reliable instrument designed specifically for assessing the creative productivity of engineering students. Through rigorous empirical validation (exploratory and confirmatory factor analyses, reliability assessments, and convergent and discriminant validity testing), ECAT takes an important first step toward addressing long-standing gaps in creativity assessment by offering a domain-specific, multi-dimensional tool.

ECAT's multi-dimensional framework, comprising fluency, originality, cognitive flexibility, and creative strengths, provides a nuanced understanding of creative performance among engineering students, highlighting critical cognitive and procedural components that are often neglected in traditional assessments. Notably, ECAT's innovative integration of flexibility, elaboration, and resistance to premature closure under the dimension of cognitive flexibility enriches existing creativity theories, emphasizing the importance of adaptability and iterative refinement in the creative engineering process.

The validation of ECAT presents a significant step toward bridging the long-standing creativity gap in engineering education, equipping educators and researchers with a robust, domain-appropriate tool to assess and cultivate creative talent systematically. By providing concrete measures of students' creative abilities and actionable insights for instructional improvement, ECAT positions itself as an essential tool not only for assessment but also for guiding curriculum development and pedagogical innovation aimed at enhancing the creative capacities of future engineering professionals.

Future research should build upon ECAT by testing its effectiveness across diverse educational contexts, employing longitudinal study designs to capture developmental trajectories of creative skills, and integrating digital and computational components to extend its applicability to emerging fields of engineering. For instance, in AI-assisted design environments, where engineers use generative tools to ideate and iterate, ECAT could serve as a valuable framework to assess students' creative contributions beyond what is generated by algorithms. Educators can be shown how to distinguish and nurture human-driven originality in partnership with machine intelligence. Similarly, in sustainability-focused engineering projects, such as designing low-cost, energy-efficient water purification systems for underserved communities, ECAT can be used to evaluate how students balance functionality with inventive use of materials, adaptability, and socially responsible design. These emerging contexts highlight ECAT's potential to guide instruction and assessment in areas where creativity, ethical awareness, and innovation are increasingly critical. Ultimately, ECAT's development signals a meaningful advancement in the pursuit of fostering creativity, innovation, and adaptability within engineering education.

# References

Acar, S., & Runco, M. A. (2019). Divergent thinking: New methods, recent research, and extended theory. *Psychology of Aesthetics, Creativity, and the Arts, 13*(2), 153–158. https://doi.org/10.1037/aca0000231

Acar, S., Tadik, H., Myers, D., Van der Schans, C., & Kaufman, J. C. (2024). Meta-analytic confirmatory factor analysis of the Kaufman Domains of Creativity Scale (K-DOCS). *Psychology of Aesthetics, Creativity, and the Arts*. Advance online publication. https://doi.org/10.1037/aca0000582

Alizamar, A., Afdal, A., Ifdil, I., & Syahputra, Y. (2019). Exploration of students' creativity based on demography. *International Journal of Innovation, Creativity and Change, 5*(1), 50-65. https://ijicc.net/images/Vol_5_iss_1_2019/alizimar_et_al_2019.pdf

Amabile, T. M. (1983). The social psychology of creativity: A componential conceptualization. *Journal of Personality and Social Psychology, 45*(2), 357–376. https://doi.org/10.1037/0022-3514.45.2.357

Amabile, T. M. (1996). *Creativity in context: Update to the social psychology of creativity*. Westview Press.

Amabile, T. M. (2012). Componential theory of creativity. In E. H. Kessler (Ed.), *Encyclopedia of Management Theory* (pp. 134–140). SAGE Publications.

Atwood, S. A., & Pretz, J. E. (2016). Creativity as a Factor in Persistence and Academic Achievement of Engineering Undergraduates. *Journal of Engineering Education (Washington, D.C.), 105*(4), 540–559. https://doi.org/10.1002/jee.20130

Baer, J. (2011). Why teachers should assume creativity is very domain specific. *International Journal of Creativity & Problem Solving, 21*(2), 57–61.

Baer, J. (2012). Domain specificity and the limits of creativity theory. *Journal of Creative Behavior, 46*(1), 16–29. https://doi.org/10.1002/jocb.002

Baer, J. (2016). *Domain specificity of creativity*. Academic Press.

Baer, J. (2017). Why you are probably more creative (and less creative) than you think. In M. Karwowski & J. C. Kaufman (Eds.), *The creative self: Effect of beliefs, self-efficacy, mindset, and identity* (pp. xx–xx). Academic Press. https://doi.org/10.1016/B978-0-12-809790-8.00014-5

Baer, J., & Kaufman, J. C. (2005). Bridging generality and specificity: The amusement park theoretical (APT) model of creativity. *Roeper Review, 27*(3), 158–163. https://doi.org/10.1080/02783190509554310

Baer, J., & Kaufman, J. C. (2017). The Amusement Park Theoretical Model of Creativity: An Attempt to Bridge the Domain-Specificity/Generality Gap. In *The Cambridge Handbook of Creativity across Domains* (pp. 8–17). Cambridge University Press. https://doi.org/10.1017/9781316274385.002

Barth, P., & Stadtmann, G. (2021). Creativity Assessment over Time: Examining the Reliability of CAT Ratings. *The Journal of Creative Behavior*, *55*(2), 396–409. https://doi.org/10.1002/jocb.462

Boateng, G. O., Neilands, T. B., Frongillo, E. A., Melgar-Quiñonez, H. R., & Young, S. L. (2018). Best Practices for Developing and Validating Scales for Health, Social, and Behavioral Research: A Primer. *Frontiers in Public Health*, *6*, 149–149. https://doi.org/10.3389/fpubh.2018.00149

Buelin-Biesecker, J. K., & Denson, C. D. (2013). Consensual Assessment: A Means of Creativity Evaluation for Engineering Graphics Education. *Web Graphics*, 64.

Carpenter, W. (2016). Engineering creativity: Toward an understanding of the relationship between perceptions of creativity in engineering design and creative performance. *International Journal of Engineering Education, 32*(5), 2016–2024.

Cattell, R. B. (1966). The scree test for the number of factors. *Multivariate Behavioral Research, 1*(2), 245-276.

Charyton, C., & Merrill, J. A. (2009). Assessing general creativity and creative engineering design in first-year engineering students. *Journal of Engineering Education, 98*(2), 145–156. https://doi.org/10.1002/j.2168-9830.2009.tb01013.x

Charyton, C., Jagacinski, R. J., & Merrill, J. A. (2008). CEDA: Assessing creativity specific to engineering design. *Psychology of Aesthetics, Creativity, and the Arts, 2*(3), 147–158. https://doi.org/10.1037/1931-3896.2.3.147

Cramond, B., Matthews-Morgan, J., Bandalos, D., & Zuo, L. (2005). A Report on the 40-Year Follow-Up of the Torrance Tests of Creative Thinking: Alive and Well in the New Millennium. *The Gifted Child Quarterly*, *49*(4), 283–291. https://doi.org/10.1177/001698620504900402

Cropley, D. H. (2015a). *Creativity in engineering: Novel solutions to complex problems*. Academic Press.

Cropley, D. H. (2015b). Promoting creativity and innovation in engineering education. *Psychology of Aesthetics, Creativity, and the Arts, 9*(2), 161–171. https://doi.org/10.1037/aca0000008

Cropley, D. H., & Cropley, A. J. (2010). Engineering creativity: A systems concept of functional creativity. In J. C. Kaufman & R. J. Sternberg (Eds.), *The Cambridge Handbook of Creativity* (pp. 301–317). Cambridge University Press.

Cropley, D. H., & Kaufman, J. C. (2012). Measuring Functional Creativity: Non-Expert Raters and the Creative Solution Diagnosis Scale. *The Journal of Creative Behavior*, *46*(2), 119–137. https://doi.org/10.1002/jocb.9

Cropley, A. J., & Maslangy, G. W. (1969). Reliability and factorial validity of the Wallach-Kogan creativity tests. *The British Journal of Psychology*, *60*(3), 395–398. https://doi.org/10.1111/j.2044-8295.1969.tb01213.x

Daly, S. R., Mosyjowski, E. A., & Seifert, C. M. (2014). Teaching creativity in engineering courses. *Journal of Engineering Education, 103*(3), 417–449. https://doi.org/10.1002/jee.20048

Denson, C. D., Buelin, J. K., Lammi, M. D., & D'Amico, S. (2015). Developing Instrumentation for Assessing Creativity in Engineering Design. *Journal of Technology Education*, 23-40. https://doi.org/10.21061/jte.v27i1.a.2

Dorrington, P., Harrison, W. J., Brown, H., Holmes, M., & Kerton, R. (2019). Step away from the CAD station: A hands-on and immersive approach to second-year teaching of mechanical engineering design. In *Proceedings of the Virtual and Augmented Reality to Enhance Learning and Teaching in Higher Education Conference 2018* (pp. 15–31).

Felder, R. M. (1988). Creativity in engineering education. *Chemical Engineering Education, 22*(3), 120–125.

Frueh, S. (2024). (2024, September). *The hallmark of engineering is creativity*. National Academies. https://www.nationalacademies.org/news/2024/09/the-hallmark-of-engineering-is-creativity

Gillier, T., Bayus, B., Marin, S.T., & Martinez-Torres, R. (2024). Creative Elaboration: When Persistence Outperforms Flexibility. *Academy of Management Journal: Proceedings.* https://doi.org/10.5465/AMPROC.2024.10225abstract

Glăveanu, V. P. (2013). Rewriting the language of creativity: The Five A's framework. *Review of General Psychology, 17*(1), 69–80. https://doi.org/10.1037/a0029528

Guilford, J. P. (1957). Creative abilities in the arts. *Psychological Review, 64*(2), 110–118. https://doi.org/10.1037/h0048280

Guilford, J. P. (1967). Creativity: Yesterday, today, and tomorrow. *Journal of Creative Behavior, 1*, 3-14.

Han, K.-S., & Marvin, C. (2002). Multiple Creativities? Investigating Domain-Specificity of Creativity in Young Children. *The Gifted Child Quarterly*, *46*(2), 98–109. https://doi.org/10.1177/001698620204600203

Hirshfield, L., & Koretsky, M. (2020). Cultivating creative thinking in engineering student teams: Can a computer-mediated virtual laboratory help? *Journal of Computer Assisted Learning, 37*(2), 587–601. https://doi.org/10.1111/jcal.12509

Hocevar, D., & Michael, W. B. (1979). The Effects of Scoring Formulas on the Discriminant Validity of Tests of Divergent Thinking. Educational and Psychological Measurement, 39(4), 917-921. https://doi.org/10.1177/001316447903900427 (Original work published 1979)

Hu, D. L., Lefton, L., & Ludovice, P. J. (2017). Humour applied to STEM education. *Systems Research and Behavioral Science, 34*(3), 216–226. https://doi.org/10.1002/sres.2406

Kaufman, J. C. (2012). *Creativity 101*. Springer Publishing Company.

Kaufman, J. C. (2016). *Creativity 101* (Second edition.). Springer Publishing Company.

Kaufman, J. C., Baer, J., Cropley, D. H., Reiter-Palmon, R., & Sinnett, S. (2013). Furious activity vs. Understanding: How much expertise is needed to evaluate creative work? *Psychology of Aesthetics, Creativity, and the Arts, 7*(4), 332–340. https://doi.org/10.1037/a0034809

Kaufman, J. C., & Beghetto, R. A. (2013). Do people recognize the four Cs? Examining layperson conceptions of creativity. *Psychology of Aesthetics, Creativity, and the Arts*, *7*(3), 229.

Kaufman, J. C., Plucker, J. A., & Baer, J. (2008). *Essentials of creativity assessment*. Wiley.

Kazerounian, K., & Foley, S. (2007). Barriers to creativity in engineering education: A study of instructors and students' perceptions. *Journal of Mechanical Design, 129*(7), 761–768.

Kershaw, T. C., Bhowmick, S., Seepersad, C. C., & Hölttä-Otto, K. (2019). A Decision Tree Based Methodology for Evaluating Creativity in Engineering Design. *Frontiers in Psychology*, *10*, 32–32. https://doi.org/10.3389/fpsyg.2019.00032

Kim, K. H. (2006). Is Creativity Unidimensional or Multidimensional? Analyses of the Torrance Tests of Creative Thinking. *Creativity Research Journal*, *18*(3), 251–259. https://doi.org/10.1207/s15326934crj1803_2

Kim, K. H. (2011). The creativity crisis: The decrease in creative thinking scores on the Torrance Tests of Creative Thinking. *Creativity Research Journal, 23*(4), 285–295.

Lucas Jr, H. C., & Goh, J. M. (2009). Disruptive technology: How Kodak missed the digital photography revolution. *The Journal of Strategic Information Systems*, *18*(1), 46-55. https://doi.org/10.1016/j.jsis.2009.01.002

Lerdal, K., Surovek, A. E., Cetin, K. S., Cetin, B., & Ahn, B. (2019). Tools for assessing the creative person, process, and product in engineering education. *Proceedings of the ASEE Annual Conference & Exposition*. https://par.nsf.gov/servlets/purl/10104575

Li, G., Chu, R., & Tang, T. (2024). Creativity Self Assessments in Design Education: A Systematic Review. *Thinking Skills and Creativity*, *52*, 101494-. https://doi.org/10.1016/j.tsc.2024.101494

Nunnally, J. C., & Bernstein, I. H. (1994). *Psychometric Theory*, 3r ed., McGraw-Hill.

OpenAI. (2025). *ChatGPT* (Mar 28 version) [Large language model]. https://chat.openai.com/chat

O'Quin, K., & Besemer, S. P. (1989). The development, reliability, and validity of the revised creative product semantic scale. *Creativity Research Journal*, *2*(4), 267–278. https://doi.org/10.1080/10400418909534323

Plucker, J. A., & Makel, M. C. (2010). Assessment of creativity. In J. C. Kaufman & R. J. Sternberg (Eds.), *The Cambridge handbook of creativity* (pp. 48-73). Cambridge University Press. https://doi.org/10.1017/CBO9780511763205.005

Pretz, J. E., & McCollum, V. A. (2014). Self-perceptions of creativity do not always reflect actual creative performance. *Psychology of Aesthetics, Creativity, and the Arts, 8*(2), 227–236. https://doi.org/10.1037/a0035597

Reiter-Palmon, R., Forthmann, B., & Barbot, B. (2019). Scoring Divergent Thinking Tests: A Review and Systematic Framework. *Psychology of Aesthetics, Creativity, and the Arts*, *13*(2), 144–152. https://doi.org/10.1037/aca0000227

Robertson, B., & Radcliffe, D. (2006). The role of software tools in influencing creative problem solving in engineering design and education. In *Proceedings of the ASME 2006 International Design Engineering Technical Conferences and Computers and Information in Engineering Conference* (Vol. 4C, 3rd Symposium on International Design and Design Education, pp. 999-1007). ASME. https://doi.org/10.1115/DETC2006-99343

Roncin, A. (2011). Enhancing confidence, competence, and connection in engineering. *Proceedings of the Canadian Engineering Education Association (CEEA)*. https://doi.org/10.24908/pceea.v0i0.3620

Runco, M. A. (2004). Everyone has creative potential. In R. J. Sternberg, E. L. Grigorenko, & J. L. Singer (Eds.), *Creativity: From potential to realization* (pp. 21–30). American Psychological Association. https://doi.org/10.1037/10692-002

Runco, M. A., & Kim, D. (2011). The four Ps of creativity: Person, product, process, and press. In M. Runco & S. Pritzker (Eds.), *Encyclopedia of Creativity* (2nd ed., pp. 534–537). Elsevier.

Runco, M. A., & Jaeger, G. J. (2012). The standard definition of creativity. *Creativity research journal*, *24*(1), 92-96. https://doi.org/10.1080/10400419.2012.650092

Sarkar, P., & Chakrabarti, A. (2011). Assessing design creativity. *Design Studies*, *32*(4), 348–383. https://doi.org/10.1016/j.destud.2011.01.002

Serice, L. G. W. (2022). *Neuroeducation and exercise: A teaching framework for multidimensional well-being and exercise sustainability* (Doctoral dissertation, Johns Hopkins University). Johns Hopkins University Institutional Repository. https://jscholarship.library.jhu.edu/handle/1774.2/XXXX

Shah, J. J., Vargas-Hernandez, N., & Smith, S. M. (2003). Metrics for measuring ideation effectiveness. *Design Studies, 24*(2), 111–134.

Silvia, P. J., Wigert, B., Reiter-Palmon, R., & Kaufman, J. C. (2012). Assessing Creativity with Self-Report Scales: A Review and Empirical Evaluation. *Psychology of Aesthetics, Creativity, and the Arts*, *6*(1), 19–34. https://doi.org/10.1037/a0024071

Sternberg, R. J. (2003). Creative thinking in the classroom. *Scandinavian Journal of Educational Research*, *47*(3), 325-338. https://doi.org/10.1080/00313830308595

Tekmen-Araci, Y., & Mann, L. (2018). Instructor approaches to creativity in engineering design education. *Journal of Mechanical Engineering Science, 233*(2), 395–402. https://doi.org/10.1177/0954406218758795

Telenko, C., Wood, K., Otto, K., Rajesh Elara, M., Foong, S., Leong Pey, K., Tan, U., Camburn, B., Moreno, D., & Frey, D. (2016). Designettes: An approach to multidisciplinary engineering design education. *Journal of Mechanical Design, 138*(2), 022001. https://doi.org/10.1115/1.4031638

Titova, O., & Sosnytska, N. (2020). The Engineer's Creative Potential Scales. *2020 IEEE Problems of Automated Electrodrive. Theory and Practice (PAEP)*, 1–4. https://doi.org/10.1109/PAEP49887.2020.9240882

Torrance, E. P. (1972). Predictive Validity of the Torrance Tests of Creative Thinking. *The Journal of Creative Behavior*, *6*(4), 236–262. https://doi.org/10.1002/j.2162-6057.1972.tb00936.x

Well, T. (2024, June 28). *The power of ambiguity*. Psychology Today. https://www.psychologytoday.com/us/blog/the-clarity/202406/the-power-of-ambiguity

Zeine, F., Jafari, N., Nami, M., & Blum, K. (2024). Awareness Integration Theory: A psychological and genetic path to self-directed neuroplasticity. *Health Sciences Review, 11*, Article 100169. https://doi.org/10.1016/j.hsr.2024.100169

Zwick, W. R., & Velicer, W. F. (1986). Comparison of five rules for determining the number of components to retain. *Psychological Bulletin, 99*(3), 432.